\documentclass[12pt]{article}

\usepackage{graphicx}
\usepackage{amsmath,amssymb}
\usepackage{bm}
\usepackage{latexsym}
\allowdisplaybreaks%
\begin{document}
\baselineskip=0.6cm
\newcommand{\EQ}{\begin{equation}}
\newcommand{\EN}{\end{equation}}
\newcommand{\EQA}{\begin{eqnarray}}
\newcommand{\EQN}{\end{eqnarray}}
\newcommand{\EQAN}{\begin{eqnarray*}}
\newcommand{\EQNN}{\end{eqnarray*}}
\newcommand{\e}{{\rm e}}
\renewcommand{\theequation}{\arabic{section}.\arabic{equation}}
\newcommand{\Tr}{{\rm Tr}}
\newcommand{\lpartial}{\buildrel \leftarrow \over \partial}
\newcommand{\rpartial}{\buildrel \rightarrow \over 
\partial}
\newcommand{\np}{{\rm :}}
\newcommand{\llbracket}{[\hspace{-1.6pt}[}
\newcommand{\rrbracket}{]\hspace{-1.6pt}]}
\newcommand{\llbracketbar}{[\hspace{-1.5pt}[ \hspace{-7pt}\smallsetminus}
\newcommand{\rrbracketbar}{]\hspace{-1.5pt}]\hspace{-10pt}\smallsetminus}
\newcommand{\vertbar}{\hspace{3pt}|\hspace{-9pt}\smallsetminus\hspace{-3pt}}
\newcommand{\slashldelimit}{\hspace{-3pt}\smallsetminus}
\newcommand{\rhdllbra}{\triangleright\hspace{-8pt}[\hspace{-1.6pt}[ \hspace{4pt}}
\newcommand{\rhdrrbra}{\, \triangleright \hspace{-5pt}\rrbracket \,}
\renewcommand{\thesection}{\arabic{section}.}
\renewcommand{\thesubsection}{\arabic{section}.\arabic{subsection}}
\makeatletter
\def\lesim{\mathrel{\mathpalette\gl@align<}}
\def\gtsim{\mathrel{\mathpalette\gl@align>}}
\def\gl@align#1#2{\lower.7ex\vbox{\baselineskip\z@skip\lineskip.2ex%
  \ialign{$\m@th#1\hfil##\hfil$\crcr#2\crcr\sim\crcr}}}
\makeatother
\makeatletter
\def\section{\@startsection{section}{1}{\z@}{-3.5ex plus -1ex minus 
 -.2ex}{2.3ex plus .2ex}{\large}} 
\def\subsection{\@startsection{subsection}{2}{\z@}{-3.25ex plus -1ex minus 
 -.2ex}{1.5ex plus .2ex}{\normalsize\it}}
\makeatother

@%
\def\thefootnote{\fnsymbol{footnote}}

\begin{flushright}
January,  2008
\end{flushright}
\vspace{0.3cm}

\begin{center}
{\Large Gravity from strings :  personal reminiscence on 
early developments}
\footnote{Invited contribution to the book
 \textit{"The Birth of String Theory"}. }


\vspace{0.4cm}

Tamiaki Yoneya


\vspace{0.3cm}

Institute of Physics, University of Tokyo \\
Komaba, Meguro-ku, Tokyo 153-8902, Japan

\vspace{1cm}
Abstract 
\end{center}
I discuss the early developments of string theory with respect to its 
 connection with gauge theory and general relativity from my own perspective. 
The period covered is mainly from 1969 to 1974, 
during which I became involved in research on dual string models 
as a graduate student. My thinking towards 
the recognition of string theory as an extended quantum theory of 
gravity is described.  Some retrospective remarks on my later works related to this subject are also given.

\section{Prologue : an encounter with the dual string model}
 I entered the graduate school of Hokkaido University, Sapporo, 
 in April, 1969. My advisor,  Akira Kanazawa 
who was an expert in dispersion-theoretic 
approach to strong interactions, 
proposed having a seminar on Regge pole theory regularly every week 
 together with a few other students. 
Comparing with various marvelous 
theories which I had learnt during undergraduate studies, however, 
the Regge pole theory was somewhat disappointing for me. 
I felt that it was too 
formal and phenomenological in its nature without much 
physical content, except for  interesting mathematical physics 
related to the notion of 
complex angular momentum. 
Looking for some more favorable  
topics, I became interested in studying the quantum field theory 
of composite particles, which, I thought, might be useful to explain 
the Regge-pole behavior from the dynamics of 
fundamental particles. I read many papers 
 related to this 
problem such as  those on compositeness criteria, on the definition of asymptotic field for a composite particle,  and especially 
on the Bethe-Salpeter 
equation. 
Although I felt that this subject was not yet 
what I really would want to pursue, I learnt much about 
quantum field theory by studying these papers. 

While still seeking subjects for my reseach,  
some senior students told me that 
there was actually a spectacular new development 
trigged off by a proposal made  a year ago by Veneziano \cite{veneziano}. 
After reading the paper of Veneziano and some others 
which extended the Veneziano amplitude to various directions, 
I gradually became convinced that this had to be the subject 
I should choose. In particular, when I was exposed to a short but 
remarkable preprint by Susskind \cite{susskind} giving a physical interpretation of the Veneziano formula in terms of vibrating strings (or `rubber band' 
in Susskind's terminology), 
I was struck by the simplicity of the idea. 
My interest on dual models was further 
strengthened by reading the paper by Nambu \cite{nambu}, 
containing 
similar discussions. In particular, I was intrigued by the concept of `master field' 
(in today's language, string field), 
which had been emphasized by Nambu. A little 
 later I also came to be fascinated by 
the very attractive world-sheet picture of Fairlie and Nielsen \cite{analogue}. 

In spite of encountering such  stunning 
ideas, my interests in dual models 
remained mostly within the  realm of more formal aspects 
of dual models during my years in 
graduate school, as was perhaps common among many other young 
students who became interested in this field in the same period.  
By the beginning of the autumn of 1969, 
I started to study 
various available preprints, 
 which were very rapidly growing. 
They included works on factorization, 
operator formalism, and unitrarization program. 

One of the papers which I was most interested in during this period 
was the one by Fubini and Veneziano \cite{fubiniveneziano}. They introduced 
an extremely elegant formulation of factorization and duality 
for general $n$-point amplitudes in terms of a 
vertex operator ${\rm :}\exp ik_{\mu}Q^{\mu}(z){\rm :}$ for tachon as 
the ground state of the dual string. 
I thought that this formalism not only was suggestive from the 
physical viewpoint of vibrating strings, but also was 
quite important in making manifest the key 
 symmetry structure, M\"{o}bius symmetry or 
SL(2, R) invariance, 
underlying the basic duality property of the amplitudes 
as had been exhibited by the Koba-Nielsen representation \cite{kobanielsen}. 
On the other hand, from the viewpoint of 
the unitarization program which had been initiated in a remarkable work by 
Kikkawa, Sakita and Virasoro (KSV) \cite{ksv}, it was a crucial step 
to construct amplitudes or vertices with the most general 
external lines corresponding to general excited states 
of strings. Attempts towards such `Reggeon' vertices were 
started by Sciuto \cite{sciuto}. During the years from 1969 to 1971, there appeared 
a large number of papers discussing this problem. 
 
 In the Japanese system of graduate school, we usually have to 
present a thesis in order to finish the first two years, 
`master course', of 
graduate study before entering the next 
stage, `doctor course', of three years.  As the subject of 
my master thesis, I decided to try, by utilizing the symmetry 
structure uncovered by the Fubini-Veneziano formalism,  
to establish a 
formalism of the Reggeon vertices 
such that the duality symmetry and factorization property were 
exhibited as manifestly as possible. 
I realized that a similar direction had been pursued in some 
papers that I became aware of, such as those by Lovelace 
\cite{lovelace} and 
Olive \cite{olive1}. I was able to achieve such a goal 
by adopting a general method proposed earlier by Shapiro 
\cite{shapiro1} 
for factorizing the generalized Veneziano amplitudes. 
I applied this method to a multiple factorization of the Fubini-Veneziano's opertorial form 
of the $n$-point amplitudes. 
The resulting general $n$-point Reggeon vertex 
enjoyed desired symmetry properties with respect to 
gauge, twisting, factorization and duality transformations. 
It also contained the previous results due to Lovelace and 
Olive as special cases. 
This small work \cite{yoneya1} became my first full paper. 
Through this work,  
I acquired a certain personal confidence in engaging myself 
further in this research field, although I was 
working alone in a way almost completely isolated from the centers 
of dual model research. 
After many years, I met C. Montonen at the Strings 2002 conference in Cambridge. 
According to him, my work had some influence on
 his work \cite
{corriganmontonen} with 
E. Corrigan in which they discussed a group theoretical 
formulation of the Reggeon vertices. In fact, I remember that 
my preprint had been cited in their paper. 
It was an unexpected surprise for me
 to have an enjoyable conversation related to
  such old works more than three decades later. 

\section{Connection of dual models to field theories}

From these early days of my studies on dual models, 
there had been one basic question which was increasingly 
occupying my mind. That was on the relationship 
between dual models and ordinary field theory.  In characterizing the Veneziano amplitude, 
it had been emphasized that the amplitude could be 
expanded into a sum over an infinite set of either 
$s$-channel poles or $t$-channel poles, but not of both. 
Adding both would amount to double-counting. 
This is of course the statement of channel duality, 
which had motivated the original proposal of the Veneziano amplitude. 
However, the unitarization program started by the KSV paper 
seemed to be an extension of the usual Feynman-graph expansion 
in ordinary field theory, in which we have to sum both 
$s$- and $t$-channel diagrams. One of my questions was 
whether it would in principle be possible to find a field theory 
from which such an extended dual Feynman-graph expansion 
would be derived, perhaps by introducing some appropriate 
interactions for Nambu's 
master field. 

Another related question 
 was on the interpretation of massless 
spinning excitations. The existence of such 
states in dual 
string models is an inevitable consequence of 
conformal symmetry associated with the 
Virasoro operators. In contrast to this, the existence of 
massless spinning fields in the framework of ordinary field theory 
is related to local gauge symmetries. 
During my master-course years, 
I became strongly interested in the 
 local gauge principle, through reading the original papers by 
Yang and Mills and by Utiyama, 
in addition to 
those related to composite particles as mentioned above. 
Because my main interest had been 
in strong interactions,  
I was naturally seeking papers aiming applications  
of the local gauge principle to strong interactions.  
In regard to later works following the above classical works, 
the paper most intriguing to me was the one by J. J. Sakurai \cite{jjs}. I was 
deeply impressed by a strong persuasive power of this work.  
I thought 
 that if the local gauge symmetry could be such a fundamental 
principle for strong interation, it should also play some role in dual models, 
and then the existence of massless spinning states in the latter 
had to be a clue.   
 
\vspace{0.2cm}
\noindent
\textit{A gauge theory interpretation of massless 
spinning states}

After submitting my master thesis in the beginning of 1971, 
I decided to pursue the second question above as my 
next theme. I attempted first to 
interpret the interaction vertices of the massless vector states 
of open strings in terms of a local
 gauge transformation for string wave 
functions. I soon realized that 
all those vertices in the Veneziano, Neveu-Schwarz \cite{dualpion} and Ramond 
\cite{ramond} models were precisely obtained  by a minimal 
substitution $p^{\mu}(\sigma) 
\rightarrow p^{\mu}(\sigma) -g A^{\mu}(x(0))\delta(\sigma)$ with $\square A^{\mu}=\partial_{\mu}A^{\mu}=0$ corresponding to 
a particular class of local-gauge transformations on the master field 
$|\Psi\rangle$, such as 
\begin{equation}
|\Psi \rangle \rightarrow e^{ig\lambda(x^{\mu}(0))}|\Psi\rangle
\label{sfieldgauge}
\end{equation}
 under an on-shell condition 
$\square \lambda(x)=0$ for 
the gauge function $\lambda(x^{\mu})$, 
where $x^{\mu}(0)$ is the string coordinate at an end point 
$\sigma=0$ of open strings. Namely, I showed that 
the vertices obtained from the Virasoro or super-Virasoro operators in this way 
coincided with those obtained by factorizing the pole residues 
at corresponding massless poles.  I interpreted this simple fact as clear 
evidence for the fact 
that these massless vector states themselves should be regarded 
as gauge bosons and that the world-sheet 
conformal symmetry was the mechanism which made the local gauge symmetry intrinsic to dual strings. In modern language, 
this formulation is of course 
equivalent to 
the vertex operator introduced as a deformation of 
the world-sheet action by background gauge field at the boundary of open-string world sheets. 

When I was pursuing this idea, I became aware of 
some prior 
works which were trying to apply local gauge principles to 
dual models, such as those  by 
Kikkawa-Sato \cite{kikkawasato} and by Manassah-Matsuda \cite{matsuda}.  
However, the standpoint common to these two works was 
entirely different from my viewpoint, in the sense that they 
were aiming toward constructing off-shell currents which 
would make possible the coupling of electromagnetic and 
weak interactions to dual strings. Hence, in their works, 
the gauge bosons coupled to the Veneziano amplitudes
 were nothing to do with the 
massless spinning states of dual strings themselves. I wrote up a  
paper on my result, emphasizing that the conformally invariant dual models of open strings should be regarded as embodying local gauge principle intrinsically, rather than interpreting the latter 
as some external structure to be artificially adjoined to the dual models. 
I submitted it to Nuovo Cim. 
Unfortunately this manuscript was rejected for the reason, as far as I remember, that the vertices discussed in it were not new. This rejection disappointed me. 
However, to my knowledge at that time, 
no one had expressed the same viewpoint as mine
 on the meaning of local gauge principle in string theory. 

A few months after this 
experience, I received a letter commenting 
on my preprint on the gauge principle in dual models 
from M. Minami, who was at the Research Institute of Mathematical Sciences
(RIMS)  in Kyoto and was actively engaged in dual models. 
To his letter, there was an attached copy of a 
preprint by Neveu-Scherk \cite{neveuscherk} on the application 
of Scherk's work \cite{scherk} on the zero-slope limit
 to the dual pion model of Neveu and 
Schwarz, making a connection 
 to gauge theory. Minami also mentioned some 
works by Nakanishi \cite{nakanishi} on a crossing-symmetric decomposition 
of the integral representation of the Veneziano-type amplitudes 
in such a way that each decomposed 
piece corresponds to a usual Feynman graph only with 
pole singularities in a single definite set of channels. 
All these works which I had not been aware of until that time
 were a big surprise to me. As for
 the rejected manuscript which I was almost trying to forget, I decided to rewrite it in a more compact form. I sent it to Prog. Theory. Phys. 
\cite{yoneya2} 
in which I added an acknowledgement to Minami. 

Minami's letter encouraged me to 
pursue my interpretation further. I immediately hit upon an idea of extending 
it to the case of closed strings, namely to the Virasoro-Shapiro    
amplitudes \cite{virasoro}\cite{shapiro2}. However, making a connection 
to gravity seemed to be a rather bizarre thing to do, since 
at that time everyone thought that the dual string 
model was aiming at the construction of a definitive theory 
for hadrons and their strong interactions. 
Actually,  during my undergraduate years, I had been quite serious about Einstein's general relativity theory 
 and had studied even some of the original 
papers.  In fact, 
Einstein had been basic impetus to me, since
 I had started out on trying to comprehend 
his book `\textit{The meaning of relativity}' during my high school day.  
However, after entering the graduate school and being influenced by the atmosphere 
of the particle physics community,  I became
 quite strongly prejudiced against 
general relativity, regarding it as a sort of a relic of the past. 
In the case of the 
vector gauge principle, I was able to free myself from similar 
prejudice by reading the Sakurai paper and others. I  hoped
 that the connection to gauge theory would be 
useful in trying to give masses to the unwanted massless states,  
in  the way $\rho$ mesons were  related to the gauge principle as 
Sakurai had advocated. But in the case of gravity, to my knowledge at that time, there had been no such attempt.  
It took another year for me to come back to this idea again. 

\vspace{0.2cm}
\noindent
{\it Extension of the Nakanishi decomposition 
and a desperate attempt toward string field theory}

Studying Nakanishi's beautiful papers on the decomposition 
of the integral representation for the $n$-point 
dual amplitudes, I gradually returned 
to my first question concerning the relationship between dual models and 
field theory. 
From the viewpoint of my question whether the dual amplitudes 
could be derived from some field theory, I thought that 
Nakanishi's way of decomposing the amplitudes had to be
one of the necessary steps. 
Then, I realized that what I had done 
in my master thesis would fit quite nicely in oder to reformulate  Nakanishi's   
result in such a way that the rule of  decomposition is 
expressed  as a generalized Feynman-like rule. That would make it  possible to extend the decomposition to arbitrary loop diagrams. 
It had been well known that a naive application of the 
operator formalism, in the case of non-planar diagrams, led to a divergence difficulty \cite{kakuthorn}, 
known as the `periodicity' problem,  of 
integrating over one and the same contribution infinitely many times. 
I was able to confirm that 
the rules I had derived on the basis of my 
previous Reggeon-vertex paper indeed resolved this difficulty as expected, because my new Feynman-like rule automatically gave 
a definite prescription for the integration region for arbitrary 
diagrams. Also this rule involving only a generalized 3-point 
vertex naturally reduced to that of 
the $\phi^3$ theory in the zero-slope limit discussed 
by Scherk. 

I was certainly excited by this result, since this strongly 
suggested that the rule I had derived might correspond precisely 
to the Feynman rule of a desired string field theory. 
When I was finishing the paper about this result in May 1972, I was staying about a month at the Yukawa Institute in Kyoto, having discussions on 
 this and related problems with Nakanishi at RIMS. 
As soon as I completed the paper \cite{yoneya3} 
on this result in Kyoto, 
I turned to attempt constructing such a formalism. 
 Seeking some hint for a new field theory underlying the dual string models, 
I studied some old works on infinite-component field theories 
and especially non-local field theories, including papers 
by Yukawa \cite{yukawa}. 
Unfortunately, however, I did not get really useful insight, 
since most of those works had been essentially restricted to free theories. 
In addition to this, 
the rule I had presented was using a very cumbersome language, which  had inherited from Nakanishi's original method and 
my own formulation of the general Reggeon vertex, and hence did not 
allow simple interpretation in terms of more intuitive world-sheet picture. 
What I was trying to do is equivalent 
to obtaining a covariant triangulation 
of the moduli space of Riemann surfaces with 
boundaries and punctures such that it corresponds to a second-quantized field theory of interacting open strings. 
More than a decade later in 1985, Witten achieved 
this goal \cite{witten}  
in his proposal of a covariant open string field theory, 
on the basis of the BRST operator formalism. 

A version of fully interacting string field theories first appeared on 
the scene in 1974, when Kaku and Kikkawa published their work 
\cite{kakukikkawa}. 
Their formulation was based on Mandelstam's 
light-cone representation \cite{mandelstam} of the 
dual amplitudes which had appeared about a year later after my 
failed attempt towards string field theory. 
The light-cone string field theory 
provided another possible, albeit \textit{non-covariant}, triangulation 
of the moduli space of Riemann surfaces. 

\vspace{0.2cm}
\noindent
{\it Connection of dual models to general relativity}

By the failure of my desperate attempt toward 
string field theory, I was quite exhausted.  To recover from a 
depressed mood which I fell into, I had to spend almost half a year from 
the summer of 1972, 
wandering about various other topics.
From about this period onward, many people were 
 departing from the dual string models. But I had some 
feeling that the dual models themselves 
contained something more to be 
uncovered, despite their immediate inapplicability to the 
physics of hadronic interactions. I was also 
influenced by a rapid resurgence of gauge field theories, 
which had been initiated by 't Hooft's famous work on the renormalizability of non-abelian gauge field theories. I was now thinking seriously about my previous 
idea on a possible connection between dual string models 
in the case of closed strings and general relativity. 
In studying the progress in gauge field theories, I was 
gradually freeing myself from the prejudice on general relativity. 
I thought that as the Yang-Mills theory had turned out to be 
useful in this way, general relativity 
might also become important for particle physics 
someday in some unexpected way. 
I also felt that it was good for me to return, now in the last stage 
of my graduate study,  to general relativity which after all had been the 
prime impetus toward theoretical physics for me. 

My thinking was something like this: One of the useful viewpoints on dual string models which had been made clear by Scherk's work was that the slope parameter 
$\alpha'$ should be treated as a fundamental constant 
as the Planck constant $h$ had been for 
quantum theory.  It would be very exciting 
if we had some kind of concrete correspondence principle between 
dual string models and general relativity, in an 
analogous way as in the case of the relation between, say,  the 
canonical commutation relation in quantum mechancis and the Poisson bracket in classical mechanics. In any case, I had first to study a 
nontrivial example of the zero-slope limit for the Virasoro-Shapiro 
amplitudes. I decided to compute a 4-point amplitude 
involving  two massive scalars and two gravitons, since the coupling 
of graviton to scalars was certainly the simplest case, as had been 
treated in most of old literatures which I had come 
across in this subject. 
All such papers of course 
only discussed the case of 4 dimensions, and I was not attempting 
to generalize them to critical dimensions, since for discussing only 
the properties of massless modes of strings without loop corrections 
 we could restrict ourselves to lower dimensions.  Consequently, the 
Newton constant was identified to be 
proportional to $g^2 \alpha'$ with $g$ being the 3-point 
coupling constant of closed strings. Staying in 
4 dimensions made possible for me 
to assign an arbitrary (real) mass to 
the scalar (tachyonic in critical dimensions) ground 
state treated as a matter field coupled with gravity, 
using additional components  of 
momentum along the extra dimensions. In the 
early summer of 1973, I decided to publish the result \cite{yoneya4} 
of such a computation first in a letter form.  

In parallel to this computation, I was pursuing 
a question on how to establish the correspondence to 
all orders with respect to graviton 
interactions in the tree approximation, or in other words 
to the tree amplitudes with arbitrary number ($=n$) of gravitons 
and with a fixed number (=2) of massive scalars. Since general relativity is a non-polynomial field theory, 
I needed  a certain general theorem 
for the purpose of establishing the connection in a 
recursive way on the basis of my explicit computation of the (2+2)-point 
amplitude. 
 I found a paper which 
was quite suitable for this aim. That was the one by 
W. Wyss \cite{wyss} who had given a very clear argument on how the 
non-polynomial Einstein Lagrangian is obtained, 
strarting from the lowest order coupling between a massive 
scalar field and a massless spin 2 field. 

The second and more ambitious question was, as alluded to already, to find an appropriate correspondence 
principle which would lead from general relativity to the dual strings 
in a more or less unique way. An ideal 
thing which I was dreaming of would be to find something analogous 
to what  had been happening in the discovery of quantum mechancis. 
I was also recalling P. Ramond's argument \cite{ramond} in his proposal of the dual  fermion model as a generalization of the Dirac equation. 
Unfortunately, however,  I was  not able to 
find such a correspondence 
principle in any satisfactory form. I could give only a modest 
discussion, which was related to a generalization of 
the equivalence principle, about the generating functional for 
the S-matrix for ($n+2$)-point scatterings:   
The S-matrix on the string side can be obtained from that of 
field theory by making a simple reinterpretation 
for the space-time energy momentum tensor of a massive scalar field. 

The third point I was thinking about was a question whether 
there could exist an extension of the Nakanishi decompostion to the 
general Virasoro-Shapiro amplitudes. That would make 
it possible to connect both sides in a more direct way 
by comparing the Feynman rules on both sides. 
I failed to arrive at really useful conclusion on this question, 
except for an expectation that in this case 
the decomposition would require most probably 
an infinite number of 
higher-point vertices as we go to the higher-point graviton 
amplitudes.\footnote{Many years later, I have 
presented some considerations 
which I had made at that time along 
this direction in a review talk \cite{yoneyakomaba} on string field theory in a workshop held at Komaba in 1986. In the late 80s, essentially the same
 problem was discussed by several groups in attempting 
to extend Witten's covariant open string field theory to 
closed strings. A systematic discussion along 
this line was given by Zwiebach
\cite{closedsft}. } 

Although  these failures somewhat disappointed me,  
I decided to write a full paper on the connection 
between the dual closed strings and general relativity. 
The title of the paper that I chose 
was `Connection of Dual Models to Electrodynamics and 
Gravidynamics',  which was sent to Prog. Theor. Phys. 
\cite{yoneya5} in the 
early autumn of 1973. 
The reason why I put also the word 
`Electrodynamics' was that I included in the paper 
an argument for the 
similar case of the open-string scattering of $n$ massless vectors and 2 massive 
scalars. That was intended for the purpose of 
emphasizing the similarity and, simultaneously, the 
contrast of the arguments 
between electrodynamics and gravidynamics, 
especially with respect to the second question above. 

I remember that, while writing this 
paper, my mind was in a somewhat perplexed 
state. One of the reasons was that I was not completely 
sure about my standpoint in interpreting the connection 
of the dual models to gravity. 
After my first seminar on my paper in our 
high-energy laboratory, a senior student (S. Araki) privately asked 
me something like this\footnote{
I cannot confirm this to Araki, since unfortunately 
he passed away in 1994. }; ``So, are you claiming that the dual model 
is more fundamental than the Einstein theory, 
replacing the latter by the  dual model?" 
I was slightly upset by this pressing question. 
As far as I remember, my answer was ``Well, as I have 
said, I could not find any definite principle,  relating the dual model to general 
relativity. If I could have provided such a deeper explanation, 
I might have a right of claiming that. All what I can say is 
simply a fact
 that we can extract general relativity from the dual model in 
this way." Rather than thinking toward 
the application of my results to the real world, 
I was mostly occupied by a desire to understand 
a deeper reason for why the theory of relativistic strings 
happens to automatically encompass general relativity, the latter being based on such deep principles as the equivalence 
principle and general covariance. I confess that 
I did not have sufficient courage to make such a bold statement, 
since I had been too strongly worried by lack of fundamental principles  
for explaining the emergence of gravity from strings. 

Furthermore, my mind was also entangled with another competing 
viewpoint on the relationship between the dual string models and 
field theory. Already in 1970, there had been 
suggestive discussions on planar `fishnet' diagrams by 
Sakita-Virasoro \cite{virasorosakita} and Nielson-Olesen 
\cite{nielsenolesen}.\footnote{
These works were precursor to the 
seminal work \cite{thooft} of 't Hooft on planar diagrams in the large $N$ limit. 
} Their elegant observation had also been quite attractive to me 
as another possible attitude towards an understanding of the relation 
between dual models and local field theories.  Moreover, in 1973, Nielsen and 
Olesen  had published an interesting pair of papers 
\cite{nielsenolesen2}\cite{nielsenolesen3} 
in which they discussed the possibility of 
embedding relativistic strings in local field theories, either 
 as a gauge theory with a 
non-linear field equation corresponding to 
the lagrangian $-\sqrt{F^2}/2\alpha'$ or as a vortex solution to 
a gauge-Higgs system, respectively. 
I became aware of these preprints just about the time when I was 
working on the connection of dual models to gravity 
on the basis of the zero-slope limit. 
These works were suggesting that the dual string 
might well be a consequence from non-linear gauge-field theories, 
resulting from their nontrivial dynamics. 
This is a view which is completely
 opposite to the derivation of field theories from the dual models 
in the zero-slope limit. 
I was asking myself,  ``Does this then imply that 
general relativity can be contained  in  gauge theory?". 

I have not been able to resolve this puzzle for many years. 
When I was invited in 1975 
from the journal `Butsuri', a periodical (in Japanese) of  
the Physical Society of Japan, to write a brief report \cite{yoneya6}
 on the recent 
development of dual models with respect to the 
relationship between the dual models and field theory, 
I began the article as: ``Since the dual resonance model 
has started out without any explicit connection to field theory, 
it has been one of the basic questions how the relationship 
between the dual model and field theory should be 
understood. There are two possible answers. 
One is that the dual string is something 
which is beyond the framework of 
ordinary local field theory, and the latter should be 
interpreted as a particular limiting case of the dual string model. 
Another possibility would be that although the dual model 
cannot be understood using the traditional perturbative methods 
of field theory, they may be properly derived from field theory
 by means of some 
non-perturbative treatments of its complicated 
non-linear dynamics. In the present report, 
I will discuss the recent developments from
 the first viewpoint, relegating 
the second to other opportunities." 

In the next year, several months after the submission of my paper, 
I received a preprint by Scherk and Schwarz \cite{scherkschwarz1} 
(see also \cite{scherkschwarz2}) in which they had 
studied the zero-slope limit of closed strings. 
In contrast to my approach of studying the coupling 
of gravitons to an artificial massive scalar state mainly for 
a methodological reason, they treated the amplitudes 
with only massless states. 
In 1975, furthermore, they made 
a bold proposal \cite{scherkschwarz3} 
that the dual string models should be 
interpreted as a unified theory of quarks and gluons. 
I was quite impressed by their works, especially 
by their clear attitude with respect to the interpretation 
of the connection of the dual models to gravity. 

At any rate, there were several reactions to my work.  In the beginning of 1974, I was invited to give a talk in a workshop 
organized by the GRG (general relativity and gravitation) group in Japan, 
and I met some of the leading Japanese physicists working 
in this field, including R. Utiyama. 
In the early summer of the same year, I received a letter from 
Y. Hara who was then staying in Europe, telling me that 
the connection of dual models to gravity was 
discussed by D. Olive in an international conference in London. 
Hara encouraged me to go to the United states for my postdoctoral 
studies. Subsequently, I received a letter from B. Sakita 
asking me whether I was interested in joining his group at the City College,  New York. 
I also met K. Kikkawa who came back to Japan  from there. 
Very stimulating conversations with him were 
most encouraging.  
Since I was actually about to be offered a faculty position  
at Hokkaido University, I joined Sakita's group, from 1976 to 1978 
during which I met A. Jevicki and S. Wadia, 
after fulfilling my teaching duty at Hokkaido for one and a half years. 
In 1980,  I moved to the Komaba campus of the 
 University of Tokyo. 
 
In these periods, my interest in dual strings 
was almost completely diminishing, because I was sparked by the 
fascinating development of gauge theories, especially 
by various new discoveries in non-perturbative aspects 
of gauge theories, such as magnetic monoples, instantons and 
lattice gauge theories. 
I became interested more and more 
in the dynamics of non-Abelian gauge theories, especially in 
the question of quark confinement, on which I was mainly working for a few 
years from the late 70s to the early 80s, though 
my papers at that time included actually 
an attempt  \cite{yozn} to derive a version of string field theory as an effective theory 
of hadronic strings directly 
from lattice Yang-Mills theory: In fact, 
I considered a generalized gauge transformation, 
 which was not so dissimilar to (\ref{sfieldgauge}),
of string field 
associated with Wilson loops  to apply it for a 
reformulation of topological structure 
of the algebra of Wilson and `t Hooft loops. 
Then from the late summer 
in 1983, I stayed at CERN for one year. During my stay 
at CERN, M. Green made a visit and gave a seminar. 
Green's talk was quite stimulating to me, especially through his enthusiasm 
for string theory. As far as I remember, he gave a review 
of his works with Schwarz, but not yet on anomaly 
cancellation. From about this time, I became gradually 
inclined again to string theory, and the impact of 
their epoch-making work \cite{greenschwarz} on the anomaly cancellation 
brought me back to this field in the mid 80s. In fact, the works 
which I published during my stay at CERN were on the 
Liouville field theory and 2 dimensional gravity, 
close cousins of 
string theory. 

\vspace{0.2cm}
\noindent
\noindent
\textit{Further related works}

Before concluding this section, let me describe 
some of my further works, which are directly related to 
my earlier papers.  After submitting the first full 
paper on gravity-string connection in 1973, I continued for sometime to think 
again 
about the possibility of establishing a Nakanishi-type 
decomposition for closed strings, since I thought that 
my discussion in connecting the closed string to general 
relativity should be improved to the level of 
a comparison between Feynman on both sides. 
In particular, I wanted to understand more clearly the mechanism 
of generating higher order interaction terms directly 
from the effect of massive excited states 
of the closed string. 
But convincing myself that 
this task was too difficult, I soon 
 turned myself to more modest problems. 
First I went back to the connection of open strings to non-Abelian gauge theory, 
and decided to establish a detailed connection at the level 
of Feynman rules  \cite{yoneya7}  on the basis of my previous work  on the general Reggeon vertex and the Nakanishi decomposition. 

Then, after a mathematical work \cite{yoneya8} on an extension 
of the Fubini-Veneziano operator formalism using superfields 
which natually arose by rewriting the 
Fairlie-Martin representation for the Neveu-Schwarz model, 
I also studied  the coupling of graviton and 
other massless closed-string states 
to fermionic open strings.  That was an extention of 
 the formalism which I had developed earlier 
in the case of massless vectors.  I showed \cite{yoneya9}\cite{yoneya10} that the vertex operators for these 
states were obtained by means of an extended local gauge principle similar to the ordinary tetrad formalism of local Lorentz 
gauge symmetry for the 
Dirac field, and discussed the 
associated geometric theory in the zero-slope limit. 
In particular in \cite{yoneya10}, I made a conjecture  on how these properties 
would be realized in full-fledged string field theories.  
Ten years later,  
I showed the validity of this conjecture, when I discussed 
the symmetry structure associated with massless 
closed string states 
in the light-cone string field theory \cite{yoneya11}\cite{yoneyadilaton}. In fact, 
this  result could have been obtained in the mid 70s 
if I had been devoting myself more seriously to the problem.  

After sending the paper on 
this subject 
to Phys. Rev. Lett.,  I received a preliminary draft 
of the astonishing paper  by Witten on a covariant open-string field theory \cite{witten}, 
in which he gave an impressive answer to the problem against which 
 I had struggled in vain 
more than 10 years ago.  But I was pleased to see that he had 
cited my latest work 
on the light-cone string field theory. 
The connection of Witten's open string 
field theory to the Nakanishi decomposition 
was noticed in
a later paper by Giddings and Martinec \cite{giddings}. 

The above idea behind the paper \cite{yoneya11} is also related to  my conjecture of a purely cubic action 
which I have proposed \cite{yoneya12} in the spring of 
1986 at the ICOBAN '86 international workshop on 
Grand Unification, held at the city of Toyama located near the site of 
the Kamiokande neutrino detector. In this workshop, there were also other talks on string theory, 
 including those by D. Gross and A. 
Strominger. Actually I had 
discussed this conjecture earlier 
in a workshop in Kyoto in the beginning of the same year 
or perhaps in the end of 1985. 
To my surprise, after only one or two months from the Toyama workshop, 
I received preprints on possible realizations of my
conjecture by Santa Barbara group \cite{cubic} and by Kyoto group \cite{pregeometry}, both mentioning my previous talk in Toyama. 

Finally, let me also mention that 
the expectation of a hidden correspondence principle, 
which I had been seeking without success in the 70s for 
characterizing the departure of string theory 
from general relativity, was basically the motivation 
for my later proposal \cite{yoneya13}\cite{yoneya15} in 1987 of an uncertainty relation for space-time. A decade later, this idea turned out to be of some relevance
 for a qualitative 
understanding  \cite{yoneya14}  of the characteristic scales of D-brane dynamics. I have written several papers \cite{yoneya16} related to this 
subject during years around the mid 1990s to the early 2000s. 

\section{Epilogue}

Recalling what I was thinking during my student years
 from 1969 to 1974, 
I find somewhat surprisingly how often my later works have been rooted 
in this period: I am still in quest for better formulation of strings and 
D-branes along various lines which 
 germinated, at least partially, from these early ideas. 
I was working almost alone, but with 
occasional kind help from other workers and, especially, 
in an environment of great progress which was  
 occuring in both 
string and gauge theories.  I was 
fortunate in being able to start  working in 
 such a dramatic period of particle physics. 

After more than three decades passed, the developments  
in string theory in the recent 10 years 
through the conception of the 
gravity-gauge correspondence (or AdS/CFT correspondence)
 are now throwing 
new light on my old perplexing question, 
the question of whether `from general relativity to gauge theory' or 
`from gauge theory to general relativity'. With respect to this 
puzzle, personally, a work \cite{yoneyaokawa} which I did with Y. Okawa in 1998 
on a derivation of the gravitational 3-body force 
for D0 branes from the D0 gauge theory 
(or M(atrix) theory \cite{bfss})  was a small revelation to me.  
This work enabled me to convince myself that general relativity may indeed be contained in the quantum dynamics of gauge field theories, when interpreted as the theory for D branes. 

In a sense, as if we are watching a revolving stage, the string theory 
came back to a scene, with which it had 
started in the early 70s, and 
is providing an entirely new perspective on gauge theory and even on hadron physics. However, we had to admit that we are still far from grasping fundamental principles behind string theory. 
Hopefully,  we will see further dramatic turns 
in the next 30 years.

\vspace{0.2cm}
I would like to thank the organizers 
for inviting me to participate in the meeting ``The birth of string theory" and in the present book.
I also would like to thank S. Wadia for useful comments 
on the manuscript.

\end{document}